\documentclass{iaus}
\usepackage{graphicx}

\title[The analysis of lithium in HD166473] 
{The analysis of Li\,{\sc i} 6708A line through the rotational period of HD166473 taking into account Paschen-Back magnetic splitting}

\author[Shavrina et al. ]   
{A.V. Shavrina$^1$, V. Khalack$^2$, Y. Glagolevskij$^3$, D. Lyashko$^4$, \\ J. Landstreet$^{5,6}$, F. Leone$^7$, M. Giarrusso$^7$
 }

\affiliation{$^1$Main Astronomical Observatory, Kyiv, Ukraine, email: {\tt shavrina@mao.kiev.ua} \\
$^2$Universit\'{e} de Moncton, Moncton, N.-B., Canada, email: {\tt khalakv@umoncton.ca} \\
$^3$Special Astrophysical Observatory, Nizhnij Arkhyz, Russia \\
$^4$Tavrian National University, Simferopol, Ukraine \\
$^5$University of Western Ontario, London, Canada \\
$^6$Armagh Observatory, Armagh, Northern Ireland –- United Kingdom\\
$^7$Universit\`{a} di Catania, Catania, Italy }

\pubyear{2013}
\volume{302}  
\pagerange{119--126}
\jname{Magnetic fields throughout stellar evolution}
\editors{A.C. Editor, B.D. Editor \& C.E. Editor, eds.}
\begin{document}

\maketitle

\begin{abstract}
The analysis of Li\,{\sc i} 6708\AA\, line was performed for 6
rotational phases distributed over the whole rotational period ($\sim$9.5 years) of HD166473. The magnetic field model was constructed based on the polarimetric measurements from \cite{Mathys_etal07}. For each observed phase the modulus of the magnetic field was also estimated from simulation of the Fe\,{\sc ii} 6147\AA, 6149\AA\, and Pr\,{\sc iii} 6706.7\AA\, line profiles taking into account Zeeman magnetic splitting. The lithium abundance in each phase was obtained from fitting the observed Li\,{\sc i} 6708\AA\, profile with the synthetic one calculated assuming Paschen-Back splitting and estimated magnetic field characteristics from Pr\,{\sc iii} 6706.7\AA\, line profile.
\keywords{Stars: chemically peculiar, stars: individual: HD166473}
\end{abstract}

\firstsection 
\section{Introduction}

High-resolution spectra of the strongly magnetic roAp star HD166473 taken for eight rotational phases spread from 0.09 to 0.97 ($P_{rot}$=3513$^d$.64) were analyzed to study the variability of the Li\,{\sc i} 6708\AA\, line profile.
The line profiles were analyzed by the method of synthetic spectra using a Kurucz model stellar atmosphere with $T_{eff}$=7750K and $\log{g}$=4.0 (\cite{Shavrina_etal06}). The magnetic splitting of Fe\,{\sc ii} 6147\AA, 6149\AA\, and Pr\,{\sc iii} 6706\AA\, lines has been calculated taking into account the Zeeman effect, while for Li\,{\sc i} 6708\AA\, line we have employed the Paschen-Back effect (\cite{KhalackLandstreet12}, \cite{Stift_etal08}).
The magnetic field measurements of \cite{Mathys_etal07} have been used to reconstruct a magnetic field configuration employing the method described by \cite{GerthGlagolevskij03}. This reconstruction results in the inclination angle of rotational axis to the line of sight $i=15^{\circ}$ and the angle $\beta=75^{\circ}$. 

\begin{table}[h]
  \begin{center}
  \caption{Abundance of chemical species at different rotational phases of
HD166473.}
  \label{tab1}
 {\scriptsize
  \begin{tabular}{|l|c|c|c|c|c|c|c|c|c|}\hline

{\bf Phase}          &  0.095 &  0.26 &  0.39 &  0.58 &  0.64 &  0.69 &  0.94 & 0.00$^1$ & solar$^2$ \\ \hline
$\log(N_{LiI}/N_H)$  & -8.20  &       & -8.42 & -8.23 & -8.20 & -8.24 & -8.23 &       & -10.95 \\
$^6Li/^7Li$          &   0.0  &       & 0.5   & 0.5   & 0.5   & 0.5   & 0.0   &	      & 0.03 \\
$\log(N_{CeII}/N_H)$ &	-7.78 &       &	-7.73 &	-7.60 &	-7.63 &	-7.64 &	-7.78 &	-7.55 &	-10.42 \\
$\log(N_{PrIII}/N_H)$&	-7.76 &       &	-7.86 &	-7.82 &	-7.80 &	-7.80 &	-7.76 &	-7.60 &	-11.28 \\
$\log(N_{NdII}/N_H)$&	-8.00 &       &	-8.10 &	-8.22 &	-8.10 &	-8.10 &	-8.30 &	-7.97 &	-10.58 \\
$\log(N_{SmII}/N_H)$ &	-8.45 &       &	-8.05 &	-7.68 &	-7.68 &	-7.75 &	-8.65 &	-8.25 &	-11.04 \\
$\log(N_{FeII}/N_H)$ &	-4.37 & -4.42 &	-4.45 &	-4.35 &	-4.35 &	-4.30 &	-4.45 &	-4.31 &	-4.50 \\ \hline
  \end{tabular}
  }
 \end{center}
\vspace{1mm}
 \scriptsize{
 {\it Notes:}\\
  $^1$Results of \cite[Gelbmann et al. (2000)]{Gelbmann_etal00},  $^2$solar data are taken from \cite[Grevesse et al. (2010)]{Grevesse_etal10}}
\end{table}

\begin{figure}[t]
\begin{center}
\includegraphics[width=1.9in,angle=-90]{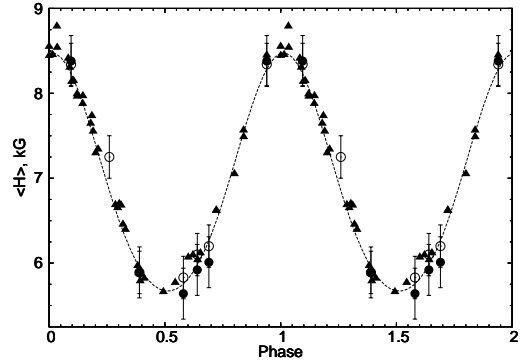}
 \caption{Variation of the mean magnetic field modulus (triangles) with rotational phase for HD166473 (\cite{Mathys_etal07}) and its approximation (dashed curve) in the framework of \cite{GerthGlagolevskij03} model. The magnetic field intensity estimated from the analysis of Pr\,{\sc iii} 6706\AA\, line (filled circles) and Fe\,{\sc ii} 6147\AA, 6149\AA\, lines (open circles) accords rather well with the Mathys' data almost for all phases. }
   \label{fig1}
\end{center}
\end{figure}

\section{Abundance analysis}

The abundances of Li and Pr obtained from the best fit of observed line profiles
are shown in the Table~\ref{tab1}. Magnetic splitting of Li\,{\sc i} line due
to Paschen-Back effect is calculated using magnetic field parameters obtained
from the modeling of the Pr\,{\sc iii} line profile. In this procedure we have
employed an idea that Li and REE lines are formed near the magnetic poles
(see \cite{Shavrina_etal01}). We have also estimated the abundances of
Ce\,{\sc ii}, Nd\,{\sc ii}  and Sm\,{\sc ii} whose lines contribute to the Li
blend (see Table~\ref{tab1}).

One can note that:
\begin{itemize}
  \item all phases show higher than "cosmic" (-8.7 dex) abundance of lithium;

  \item some differences in the abundance of Li\,{\sc i}, the isotopic ratio $^6Li/^7Li$, and REE abundances for different phases,
and rather different values of the magnetic field strength obtained from the
Pr\,{\sc iii} 6706.7\AA\, profile, can be explained by the different location
of lithium and REE spots,
and by their different stratification
with optical depth;
  \item the value of $^6Li/^7Li$ ratio can be consider as a parameter(measure) of
model Li line profile disagreement with observed one;

  \item additional spectra of this very interesting star (high resolution and
high S/N) are required to thoroughly cover the whole rotational period.

\end{itemize}

\end{document}